\documentclass[preprint,prl]{revtex4}

\usepackage{graphicx}
\usepackage{dcolumn}
\usepackage{bm}
\usepackage{float}
\usepackage{amsmath}
\usepackage{graphicx}
\usepackage{dcolumn}
\usepackage{bm}
\usepackage{hyperref}
\usepackage{cleveref}
\newcommand{\be}{\begin{equation}}
\newcommand{\ee}{\end{equation}}
\newcommand{\bea}{\begin{eqnarray}}
\newcommand{\eea}{\end{eqnarray}}
\begin{document}

%
%
%

\title{Bipartite and Tripartite Entanglement for Three Damped Driven Qubits}


\author{William Konyk}
\affiliation{Department of Physics, University of Arkansas, Fayetteville AR 72701.}
\author{Ethan Stanifer}
\affiliation{Department of Physics, Syracuse University, Syracuse NY 13201}
\author{Habtom Woldekristos}
\author{James Clemens}
\author{Perry Rice}
\affiliation{Macklin Quantum Information Sciences \\Department of Physics, Miami University, Oxford,
Ohio 45056}

\date{\today}

\begin{abstract}
We investigate bipartite and tripartite entanglement in an open quantum system, specifically three qubits, all of which are damped, and one of which is driven. We adapt a systematic approach in calculating the entanglement of various bipartite splits usinga generalized concurrence as an indicator  of entanglement.  Our calculations are based on a direct detection scheme  that is a particular unravelling of the density matrix. This system has a collective dipole-dipole energy shift that couples the atoms and the dissipation is via partially collective spontaneous emission described by the Lehmberg-Agarwal master equation.Our results are unravelling dependent, but apply to applications of entanglement based on direct detection. We also calculate the three-way tangle or residual entanglement for this system. We present calculations for a variety of driving and damping rates, and examine what decay rate is adequate for the system to be reduced to two qubits with a readout port.  We also consider a specific model of three atoms located at particular positions in free space.  
\end{abstract}

\pacs{03.65.Ud, 42.50.Nn}

  \maketitle
\section{Introduction}
Bipartite and tripartite entanglement and collective atomic interactions are topics of interest for fundamental reasons and because of applications in quantum information processing protocols   In the present work we consider the tripartite entanglement and the entanglement of various bipartite splits of a system of three two-level atoms with one atom driven on resonance with an external field and all three atoms damped by their individual (collective) coupling to the vacuum electromagnetic field.

Over the last two decades, studies into the foundations of quantum mechanics have shown us that entanglement is an inherently quantum mechanical property, which may be used as a resource for tasks such as factoring a number that is the product of two large prime numbers, search protocols, and quantum teleportation \cite{ent1,ent2}. The measure of entanglement is well defined primarily in the case of two interacting qubits in a pure state. Consider a two qubit state
\be
|\psi\rangle=C_{00}|00\rangle+C_{01}|01\rangle+C_{10}|10\rangle+C_{11}|11\rangle
\ee
If the system is in a product state
\be
|\Psi\rangle=\left(A_0|0\rangle+A_1|1\rangle\right)\left(B_0|0\rangle+B_1|1\rangle\right)
\ee

Then the coefficients will satisfy
\be
{\cal E}=C_{00}C_{11}-C_{01}C_{10}=0
\ee
 If this relation is {\it not} satisfied then the state is entangled. The concurrence is a measure of entanglement for two qubits and is ${\cal C}=\sqrt{2{\cal E}}$ \cite{con}.

 For three qubits, they can be entangled in different ways,  a Greenburger-Horne-Zeilinger state\cite{ghz}, a W state, or entanglement solely between various bipartite splits of the system (i.e.~particle 1 entangled with 2 and 3 etc.) \cite{tri}. If we have two coupled qubits, that interact with their environment, being pumped with energy or dissipative processes, we have a similar problem. Essentially we have three systems, qubit 1, qubit 2, and the environment.  Just considering the two qubits by tracing, or averaging over the environmental degrees of freedom, we have two qubits described by a mixed state instead of a pure state. There are many types of entanglement measures and entanglement witnesses \cite{m1}.

In this paper we use a pragmatic measure of entanglement introduced by Nha and Carmichael \cite{Nha} that utilizes quantum trajectory theory. Open quantum systems are often described by a reduced density matrix that is obtained by tracing, or averaging, over environmental degrees of freedom. Typically one uses a Born-Markov approximation to derive the master equation for the reduced density matrix. Recall that in quantum trajectory theory, the density matrix is unravelled \cite{traj1,traj2,traj3}
\be
\dot{\rho}={\cal L}\rho=\left({\cal L-S}\right)\rho+{\cal S}\rho
\ee
where ${\cal L}$ is the Liouvillian operator defined by the master equation, and ${\cal S}$ can be any operator whatsoever. For a given choice of ${\cal S}$, the ${\cal L-S}$ part of the evolution can be described by an effective Schroedinger equation, with a non-Hermitian Hamiltonian evolving $|\psi_c\rangle$. This is often referred to as a quantum trajectory. This evolution is punctuated by the application of the ${\cal S}$ operator at times randomly chosen from a distribution that is determined by the current state of the system, which we refer to as a jump. At every time step of course the trajectory is normalized, as a jump or non-Hermitian evolution both result in nonunitary evolution. To fully recover the statistical information about the system, one then averages over  a set of trajectories. Obviously a different choice for ${\cal S}$ leads to a different unravelling and a different set of trajectories, so this is not the same as a wave function for a pure state. One common choice for ${\cal S}$ is that of direct detection, where for example we monitor the spontaneous emission of an atom via a $4\pi $ detector with perfect efficiency. This yields quantum trajectories conditioned on the measurement record. Another common choice is that for homodyne detection. There the number of jumps is very high over the characteristic dissipative/driving rates of the system and one coarse grains the resulting equation which results in a nonlinear Schroedinger equation for evolution. A nice description in terms of measurement theory has been given by Wiseman \cite{wiseman} and Jacobs and Steck \cite{jacsteck}.  Nha and Carmichael proposed applying pure state entanglement measures to these trajectories, yielding a functional definition of entanglement for open systems. Obviously, the amount of entanglement obtained is {\it different} for different choices of ${\cal S}$. They explicitly demonstrated this for the choice of monitoring their optical system via homodyne and direct detection. In this paper we will pursue this method of examining entanglement of open systems. It has been shown that using trajectories, for a given choice of unravelling, the behavior of many entanglement measures behave similarly under change of parameters. \cite{trajwork} This method has also been utilized recently for cavity QED. \cite{robleno} For tripartite partitions of the system we use the three-way tangle to characterize entanglement. This measure is zero for W states and unity for GHZ states \cite{3tangle}.

\section{Methods}

The master equation describing the evolution of the atomic density operator with the electric-dipole, rotating-wave, and Born-Markov approximations in the interaction picture is given by:\cite{Russ,Lehmberg,Agarwal,Clemens}
\begin{equation} \dot{\rho} = iY [ (\sigma_{1-}- \sigma_{1+}),\rho ]- i \sum_{i \neq j = 1}^{3} \Delta_{ij} [ \sigma_{i+} \sigma_{j-} , \rho ]  +\frac{1}{2} \sum_{i,j=1}^{3} \gamma_{ij}(2 \sigma_{j-} \rho \sigma_{i+}- \sigma_{i+} \sigma_{j-} \rho-\rho \sigma_{i+}\sigma_{j-})
\end{equation}
where $Y$ is the driving field strength, $\sigma_{j-}$ is the Pauli lowering operator for the $j$th atom,
\begin{equation}\label{eq:Delta}\Delta_{ij}=\gamma \frac{3}{4}[-(1-(\hat{d}\cdot \hat r_{ij})^{2})\frac{cos(\xi_{ij})}{\xi_{ij}}+(1-3(\hat{d}\cdot \hat r_{ij})^{2})(\frac{sin(\xi_{ij})}{\xi_{ij}^{2}}+\frac{cos(\xi_{ij})}{\xi_{ij}^{3}})],
\end{equation}
\begin{equation}\gamma_{ij}=\gamma \frac{3}{2}[-(1-(\hat{d}\cdot \hat r_{ij})^{2})\frac{sin(\xi_{ij})}{\xi_{ij}}+(1-3(\hat{d}\cdot \hat r_{ij})^{2})(\frac{cos(\xi_{ij})}{\xi_{ij}^{2}}-\frac{sin(\xi_{ij})}{\xi_{ij}^{3}})], 
\end{equation}
$\gamma$ is the spontaneous emission rate for each atom, and
\begin{equation} \xi_{ij}\equiv \frac{2\pi r_{ij}}{\lambda}, \quad r_{ij} \equiv |\vec{r_{i}}-\vec{r_{j}}|. 
\end{equation}
The $\Delta_{ij}$ terms describe the dipole-dipole coupling of the atoms and the $\gamma_{ij}$ terms describe collective spontaneous emission.  For simplicity we assume the polarization of the driving laser is normal to the plane defined by the location of the three atoms so that $\hat d\cdot \hat r_{ij} = 0$.

A system of three arbitrary qubits (which may be two-level atoms) which may be entangled to each other is considered as shown in the figure \ref{fig:threequbit} below.
\begin{figure}[htp]
	\centering
		\includegraphics [height=4in, width =4in]{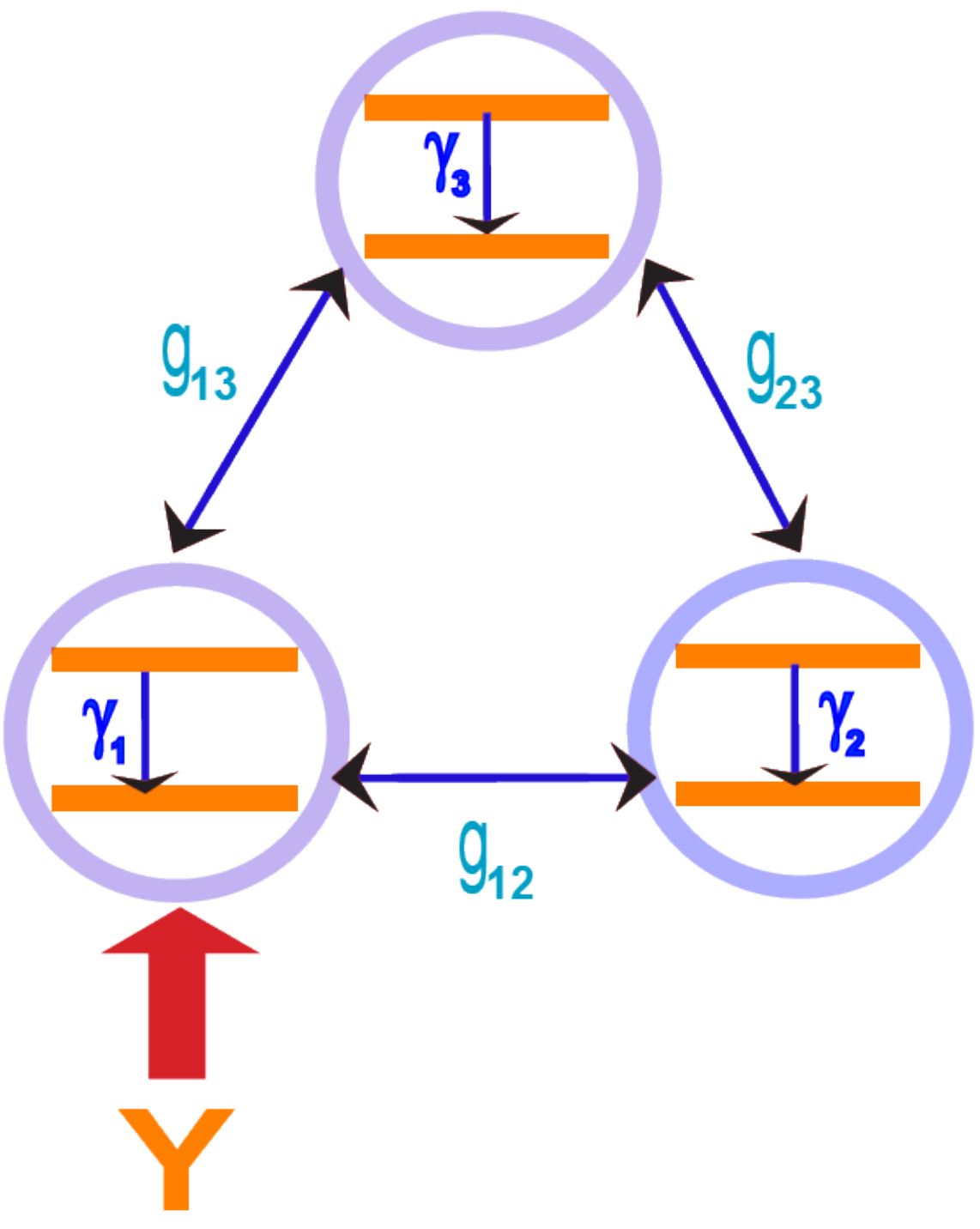}
	 \caption{Three qubits model}
	 \label{fig:threequbit}
\end{figure}
A general pure state for the three-qubit system is given by
\begin{eqnarray}
\label{eq:threequbit}
|\psi \rangle & = & C_{000}|000\rangle + C_{001}|001\rangle + C_{010}|010\rangle + C_{011}|011\rangle + C_{100}|100\rangle \\
              & &+ C_{101}|101\rangle + C_{110}|110\rangle + C_{111}|111\rangle
\end{eqnarray} 
and the Hamiltonian of the system is
\begin{eqnarray}
\label{eq:threeqhamil}
H &=& \hbar g_{12}(\sigma_{1+}\sigma_{2-} + \sigma_{2+}\sigma_{1-}) + \hbar g_{13}(\sigma_{1+}\sigma_{3-} + \sigma_{3+}\sigma_{1-})  
+ \hbar g_{23}(\sigma_{2+}\sigma_{3-} + \sigma_{3+}\sigma_{2-}) \\ 
& & + \imath \hbar Y(\sigma_{1+} - \sigma_{1-}) - \imath \hbar \gamma_1 \sigma_{1+} \sigma_{1-} - \imath \hbar \gamma_2 \sigma_{2+} \sigma_{2-} - \imath \hbar \gamma_3 \sigma_{3+} \sigma_{3-}.
\end{eqnarray}
where we include non-Hermitian terms to account for the effects of the dissipation in time intervals where there is no jump. The state of the system evolves according to the Shroedinger equation  and we get the following equations of the probability amplitude rates:

\begin{eqnarray}
\label{eq:probAmp3q1}
\dot{C}_{000} &=& - YC_{100} \\
\dot{C}_{001} &=& - \imath g_{13}C_{100} - \imath g_{23}C_{010} - YC_{101} - \gamma_3C_{001}\\
\label{eq:probAmp3q2}
\dot{C}_{010} &=& - \imath g_{12}C_{100} - \imath g_{23}C_{001} - YC_{110} - \gamma_2C_{010}\\
\label{eq:probAmp3q3}
\dot{C}_{011} &=& - \imath g_{12}C_{101} - \imath g_{13}C_{110} - YC_{111} - (\gamma_2 + \gamma_3)C_{011}\\
\label{eq:probAmp3q4}
\dot{C}_{100} &=& - \imath g_{12}C_{010} - \imath g_{13}C_{001} + YC_{000} - \gamma_1C_{100}\\
\label{eq:probAmp3q5}
\dot{C}_{101} &=& - \imath g_{12}C_{011} - \imath g_{23}C_{110} + YC_{001} - (\gamma_1 + \gamma_3)C_{101}\\
\label{eq:probAmp3q6}
\dot{C}_{110} &=& \imath g_{13}C_{011} - \imath g_{23}C_{101} + YC_{010} - (\gamma_1 + \gamma_2)C_{110}\\
\label{eq:probAmp3q7}
\dot{C}_{111} &=& YC_{011} - (\gamma_1 + \gamma_2 + \gamma_3)C_{111}
\label{eq:probAmp3q8}
\end{eqnarray}

The state initially prepared is the ground state of the system. That is
\begin{equation}
\label{eq:grdst}
|\psi(0)\rangle = |000\rangle
\end{equation}
where $C_{000} = 1$ and all the other coefficients are zero. The state then starts to evolve by equations (\ref{eq:probAmp3q1} - \ref{eq:probAmp3q8}) followed by collapses or jumps generated stochastically by the jump operators $\sqrt{\gamma_{1}}\sigma_{1-}, \sqrt{\gamma_{2}}\sigma_{2-}$ and $\sqrt{\gamma_{3}}\sigma_{3-}$. The three jump probabilities are
\begin{eqnarray}
p_{1}(t) = \left(\left|C_{100}\right|^{2} + \left|C_{101}\right|^{2} + \left|C_{110}\right|^{2} + \left|C_{111}\right|^{2}\right)\gamma_{1}\Delta t \\
\label{eq:prob1}
p_{2}(t) = \left(\left|C_{010}\right|^{2} + \left|C_{011}\right|^{2} + \left|C_{110}\right|^{2} + \left|C_{111}\right|^{2}\right)\gamma_{2}\Delta t \\
\label{eq:prob2}
p_{3}(t) = \left(\left|C_{001}\right|^{2} + \left|C_{011}\right|^{2} + \left|C_{101}\right|^{2} + \left|C_{111}\right|^{2}\right)\gamma_{3}\Delta t 
\label{eq:prob3}
\end{eqnarray}
Each trajectory is simulated keeping a record of the state of the system over a total time of 10 lifetimes and the state of the system evolves by equations (\ref{eq:probAmp3q1} - \ref{eq:probAmp3q8}) or the evolution is interrupted by collapses each random time by one of the three jump operators. A random number is generated and compared with the total jump probability to know whether a jump is detected or not and another random number is thrown to exactly know which qubit jumps. 

We calculate the concurrences $C_{1,23}, C_{2,13}$ and $C_{3,12}$ based on the direct detection scheme over many trajectories until they get to a steady state. For example, the concurrence between qubit 1 and the pair qubits (2,3) is
\begin{equation}
\label{eq:c123}
C_{1,23} = \sqrt{2(1-tr(\rho_1^2))}
\end{equation}
where $\rho_1 = tr_{23}(|\psi\rangle\langle\psi|)$, i.e.~tracing over qubit 1 of the density matrix of the system gives us the entanglement between that qubit and the pair of qubits 2 and 3. And similarly
\begin{eqnarray}
	C_{2,13} &=& \sqrt{2(1-tr(\rho_2^2))}\\
	C_{3,12} &=& \sqrt{2(1-tr(\rho_3^2))}.
	\label{eq:c2n3}
\end{eqnarray}
For a pure state this measure of entanglement works because if the state can be factorized then tracing over one subsystem leave the other in a pure state whereas if the state is entangled then tracing over one subsystem leaves the other in a mixed state.

\subsection{Atoms in free space}
We consider an explicit model consisting of three atoms located at fixed positions in free space.  For simplicity, the three atoms are placed on a plane while the polarization of the driving light is normal to the plane. Since the length scale and time scale are set by the fundamental wavelength, $\lambda$, and atomic lifetime, $\gamma$, Fig.~\ref{fig:coupling} shows the plot of the position dependent coupling $\Delta_{ij}$ and collective spontaneous emission parameter $\gamma_{ij}$.

The Quantum Toolbox in Python, QuTiP, has a function which performs a Monte-Carlo simulation for numerous trajectories given the Hamiltonian and Jump operators. The $\Delta_{ij}$ is the coupling strength between atoms $i$ and $j$ which is easily placed in the Hamiltonian, with the addition of $\Delta_{ij}\sigma_{i+}\sigma_{j-}$. The Hamiltonian in general for this simulation is given by:
\begin{equation} H=i\hbar Y (\sigma_{1-}-\sigma_{1+})+\sum^{3}_{j=1}\sum^{3}_{i=1}\Delta_{ij}\sigma_{i+}\sigma_{j-}
\end{equation}
where $Y$ is the driving strength and $\Delta_{ij}$ is calculated using Eq.~\ref{eq:Delta}.
The normalized eigenvectors of the matrix $\gamma_{ij}$ can be used to construct the jump operators, weighed by the eigenvalues \cite{Clemens}.

\begin{figure}[htp]
\centering
		\includegraphics[width=7in]{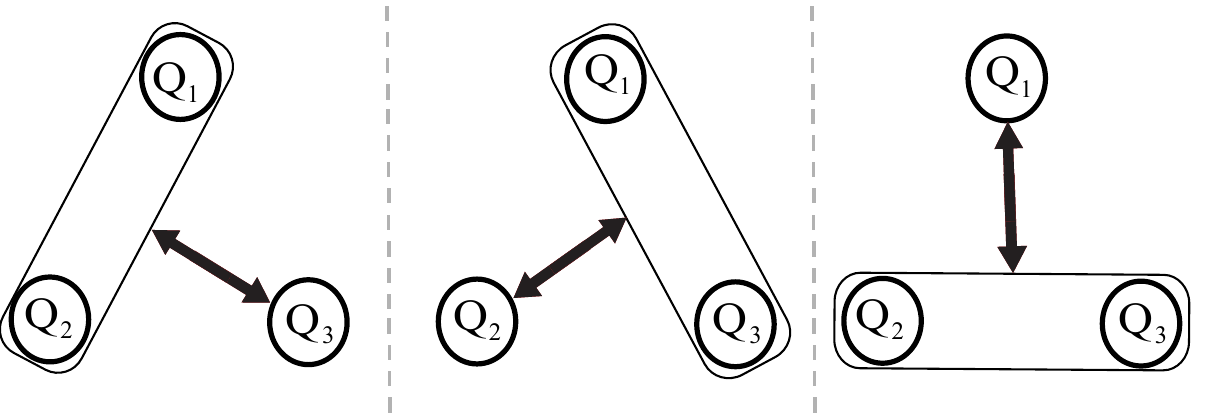}
	 \caption{Bipartite splits between the three qubits Q1,Q2 and Q3.}
	 \label{fig:splits}
\end{figure}

These three qubits 1,2 and 3 may be entangled with each other based on the values of the parameters in equation (\ref{eq:threeqhamil}). We calculated the entanglement between various bipartite splits specifically the concurrence between the state of one qubit and the state of the other two qubits. The bipartite splits are shown in Fig.~\ref{fig:splits}. 

Starting from the conditional pure state generated by the quantum trajectory algorithm the reduced density operator on atom $i$ can be defined by
\begin{equation} 
\rho_{i}=Tr_{jk}[|\Psi\rangle\langle\Psi|]
\end{equation}
tracing over qubits $j$ and $k$ of the density matrix of the system. We calculate the concurrence for the bipartite split between atom $i$ and the pair of atoms $j$ and $k$ 
\begin{equation}
C_{i,jk}= \sqrt{2-2Tr[\rho_{i}^{2}]}
\end{equation}
where this measure of entanglement is conditioned on a direct photo detection of the light scattered by the atoms\cite{Nha}.  To quantify tripartite entanglement we use the residual entanglement or the three tangle  \cite{3tangle} which is given by $\tau_{123}=C_{1,23}^2-C_{1,2}^2-C_{1,3}^2$.

We numerically solve the master equation for the system by means of a quantum trajectory simulation implemented using the Quantum Toolbox in Python (QuTiP) \cite{Johansson-1760-2012, Johansson-1234-2013}.  In order to simulate the collective spontaneous emission we use construct the jump operators using eigenvectors of the $\gamma_{ij}$ matrix to weight the contribution of each atom to the emission \cite{Clemens}.

In Fig. (\ref{fig:triangles})we show three  configurations we consider initially. In these configurations, decay is dominated by single atom emission, collective emission, and a mixture. Fig (4.) shows the collective damping and coupling terms between two atoms as a function of their separation.

\section{Results} 
\begin{figure}
\centering
\includegraphics[width=5in]{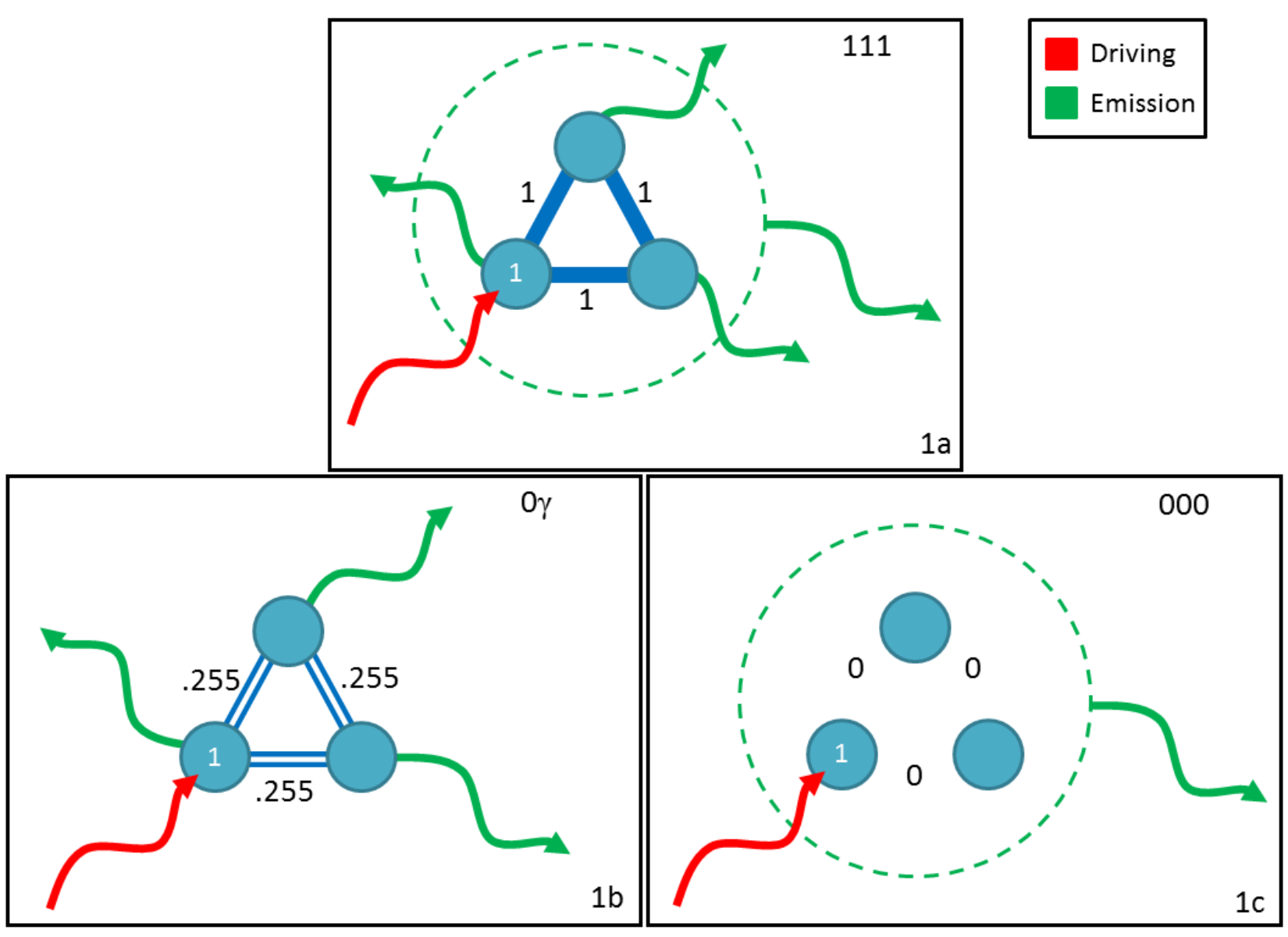}
\caption{a) The coupling strengths are $1$ and emissions are partially collective and partially not collective. b) The coupling strengths are $.255$ and emissions are totally distinguishable. c) The coupling strengths are $0$ and the interesting emissions are collective.}
\end{figure}

\begin{figure}
\center{\includegraphics[width=4in]{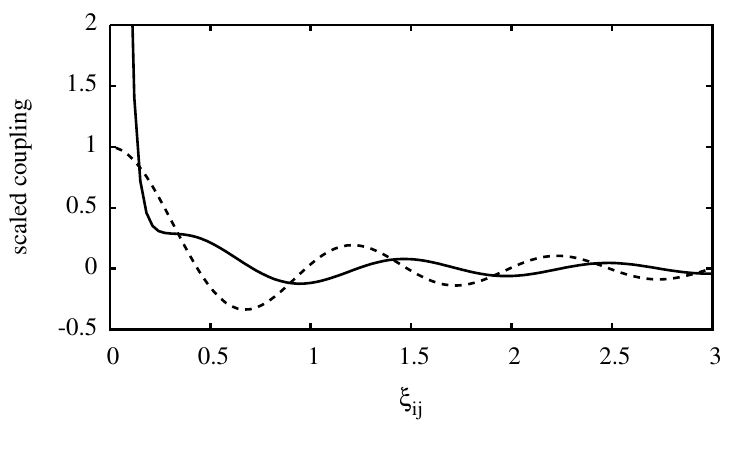}}
\caption{\label{fig:coupling}The scaled off diagonal coupling rates $\Delta_{ij}/\gamma$ (solid) and $\gamma_{ij}/\gamma$ (dashed) versus scaled spacing $\xi_{ij} = 2\pi r_{ij}/\lambda_0$.}
\end{figure}

\begin{figure}
\centering
\includegraphics[width=2in]{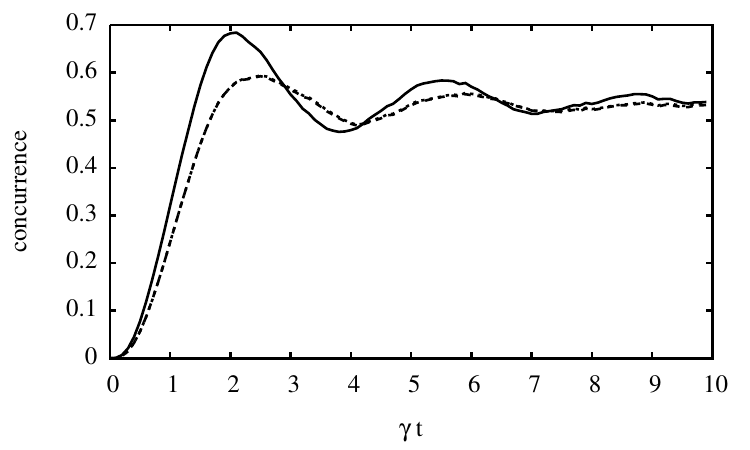}\includegraphics[width=2in]{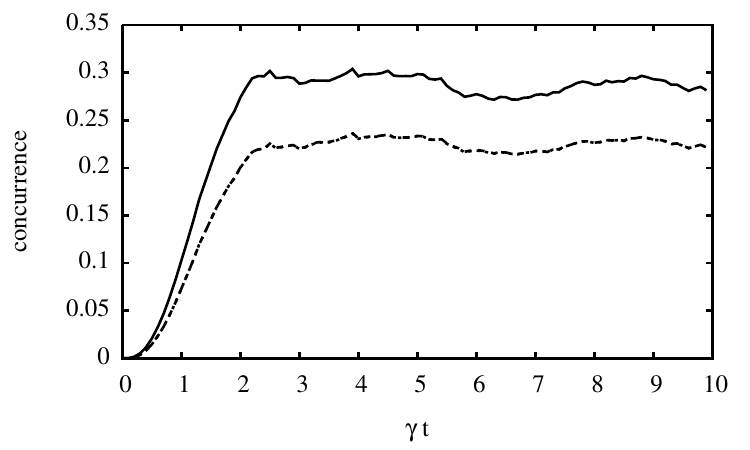}\includegraphics[width=2in]{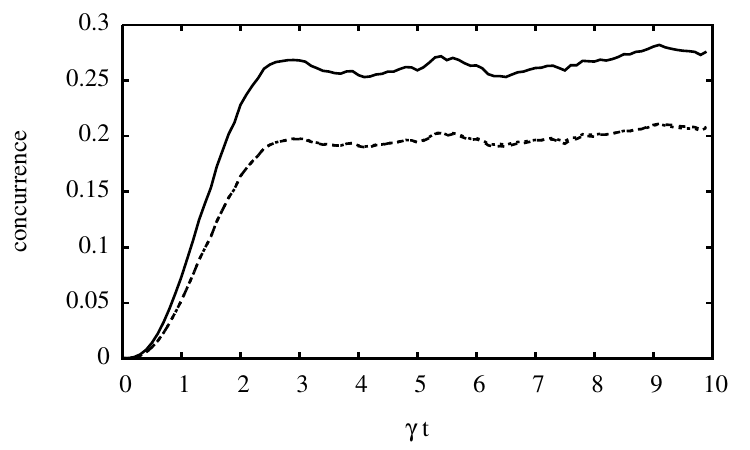}
\caption{$C_1$ (solid), $C_2$ (dashed), $C_3$ (dotted) for the equilateral triangle for which (a) the coupling is 1, (b) the emission is fully independent, and (c) there is no dipole-dipole coupling. \label{fig:triangles}}
\end{figure}

We considered three particular choices for the arrangement of the atoms on the vertices of an equilateral triangle with the dipole moment perpendicular to the plane of the triangle
The first case was a spacing chosen so that all dipole-dipole couplings are $\Delta_{ij} = \gamma$ which is accomplished by an equilateral triangle with a side of length $0.135\lambda_0$. This case served as a mixed case since it has relatively strong coupling and has somewhat collective emission. The resulting entanglement, shown in Fig.~\ref{fig:triangles}(a), shows the entanglement calculated from having a coupling strength of 1 and a spontaneous emission matrix which is not the identity. However, the true source of this entanglement is not fully exposed in this graph.

The second case was chosen so that the emission is fully independent; this is achieved with a spacing of $0.44\lambda_0$ (see Fig.~\ref{fig:coupling}). This is the closest distance at which the off diagonal entries of $\gamma_{ij}$ are zero which results in dipole-dipole couplings $\Delta_{ij} = .255\gamma$.  The concurrence for this case is shown in Fig.~\ref{fig:triangles}(b).

The last case is the closest distance for which there is no dipole-dipole coupling; this is a spacing of $0.713\lambda_0$. This position yields an off diagonal entry of $\gamma_{ij} = -.32 \gamma$ which creates combined jump operators: $J_{1}=.35 \gamma \left( \begin{matrix} -.577, -.577, -.577 \end{matrix} \right)$, $J_{2}=1.32 \gamma \left( \begin{matrix} .535,  -.802 ,.267 \end{matrix} \right)$, and $J_{3}=1.32 \gamma \left( \begin{matrix} .617\\ .154\\ -.772 \end{matrix} \right)$. The intuition is that since there is no dipole-dipole coupling there should be no concurrence; however, this case, yields almost as much concurrence, shown in Fig.~\ref{fig:triangles}(c), as the second case.  Nevertheless, the partial collective emissions of this case yields concurrence. At first glance, it would seem that the nondriven atoms should stay in the ground state since they are not directly driven and there seems to be no reason that spontaneous emission from the first atom should not cause the other atoms to become excited.  However, this is not the case.

To illustrate this we consider the two atom case;  the driving takes system from $|gg\rangle$ to $|eg\rangle$. We can define linear superposition states $|-\rangle$ and  $|+\rangle$ as
\begin{equation}
|\pm\rangle=\frac{\sqrt{2}}{2}(|eg\rangle\pm|ge\rangle)
\end{equation}
which can be inverted to find
\begin{equation}
|eg\rangle=\frac{\sqrt{2}}{2}(|+\rangle+|-\rangle).
\end{equation}
As this evolves over time, the $|-\rangle$ state does not decay since collective spontaneous emission is symmetric but $|+\rangle$ will decay. Therefore, as $|eg\rangle$ evolves in time it develops a nonzero overlap with $|ge\rangle$ and therefore the concurrence is nonzero.  In the development of the Lehmberg-Agarwal superradiance master equation this process appears as an explicit coupling between the first atom and the field, then a subsequent coupling between the field and the second atom.  The field is adiabatically eliminated and we are left with an effective direct atom-atom coupling.

\begin{figure}
\begin{minipage}[t]{0.49\columnwidth}
\centering
\includegraphics[width=1\textwidth]{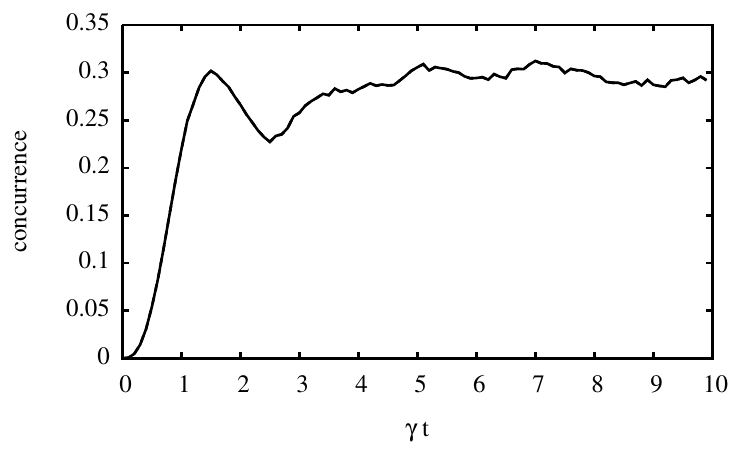}
\par\centering
(a)
\end{minipage}
\hfill
\begin{minipage}[t]{0.49\columnwidth}
\centering
\includegraphics[width=1\textwidth]{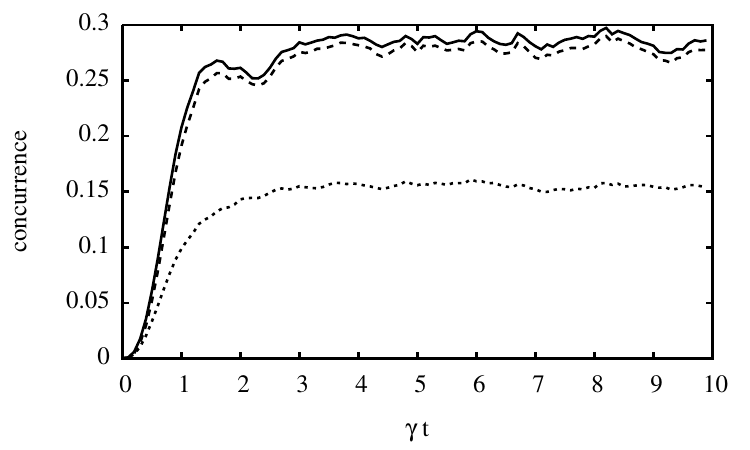}
\par\centering
(b)
\end{minipage}
\hfill
\begin{minipage}[t]{0.49\columnwidth}
\centering
\includegraphics[width=1\textwidth]{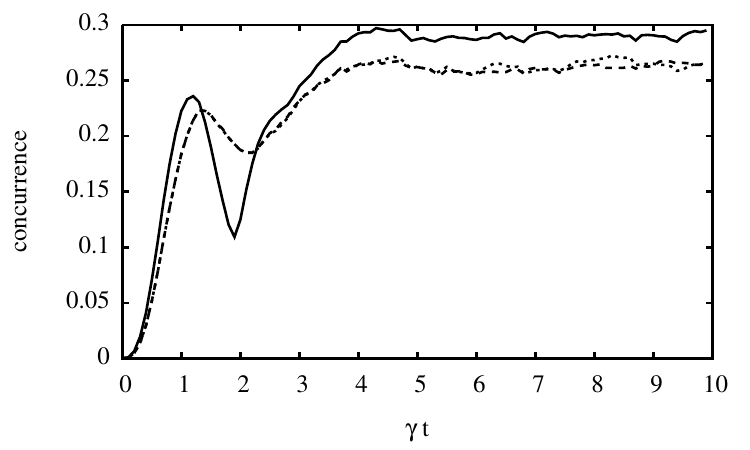}
\par\centering
(c)
\end{minipage}
\hfill
\begin{minipage}[t]{0.49\columnwidth}
\centering
\includegraphics[width=1\textwidth]{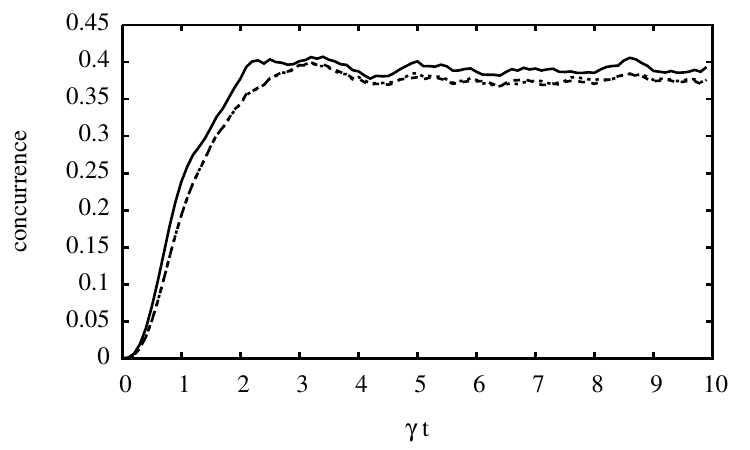}
\par\centering
(d)
\end{minipage}
\caption{A plot of concurrences, $C_{1,23}$ (solid), $C_{2,13}$ (dashed) and $C_{3,12}$ (dotted), calculated with knowledge of which qubit jumps. $Y = 1$.}
\label{fig:con11}
\end{figure}

We also examine the entanglement when we assume the atomic positions are such that we can have only collective, or single atom decay, with a coupling of $g$ in Fig.~\ref{fig:con11}(a). This reaches a steady state value after about 6 lifetimes. Two of the bipartite splits have equal entanglement, the ones where the uncoupled qubit is paired with one of the others, the entanglement between the uncoupled one and the coupled pair is of course zero. In Fig.~\ref{fig:con11}(b) we set $g_{12}=0$, and find that the entanglement between bipartite splits that was nonzero before is now slightly less and are now unequal. In Fig.~\ref{fig:con11}(c) we set $g_{23}=0 $ with the other two couplings equal. Here we find that after a transient oscillation the steady state is similar to Fig.~\ref{fig:con11}(b) with $C_2$ and $C_3$ essentially identical, excepting statistical fluctuations. Finally in Fig.~\ref{fig:con11}(d), we have all three couplings nonzero and equal. Here again we have a symmetry and two of the bipartite splits have equal entanglement with the maximum entanglement between the driven qubit and the pair of undriven qubits.  Overall the entanglement is slightly larger for this case.

\begin{figure}
\begin{minipage}[t]{0.49\columnwidth}
\centering
\includegraphics[width=1\textwidth]{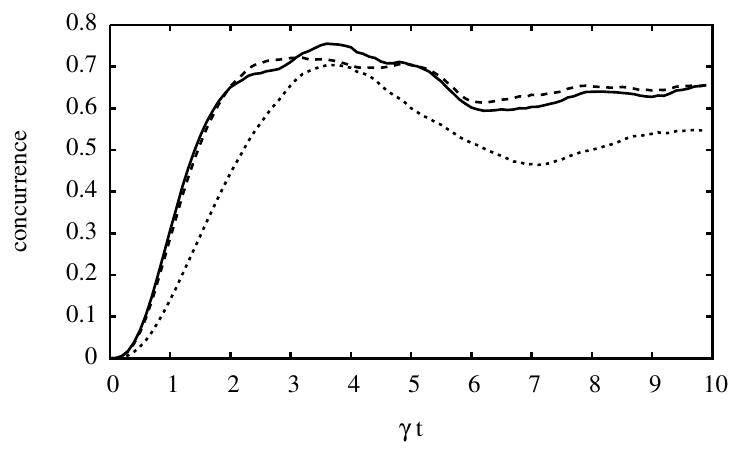}
\par\centering
(a)
\end{minipage}
\hfill
\begin{minipage}[t]{0.49\columnwidth}
\centering
\includegraphics[width=1\textwidth]{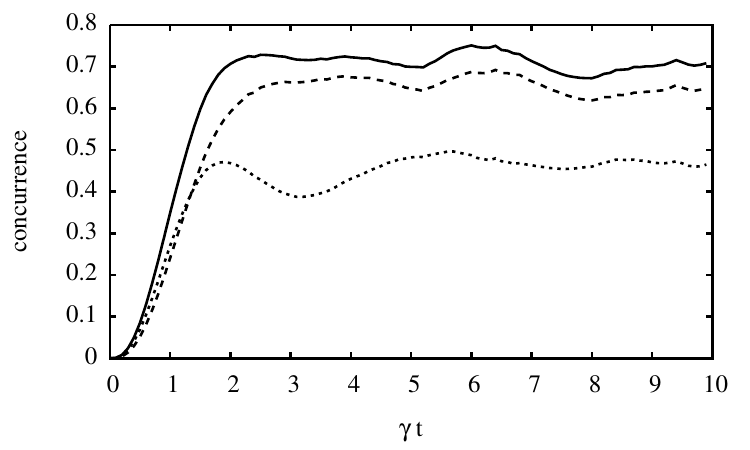}
\par\centering
(b)
\end{minipage}
\hfill
\begin{minipage}[t]{0.49\columnwidth}
\centering
\includegraphics[width=1\textwidth]{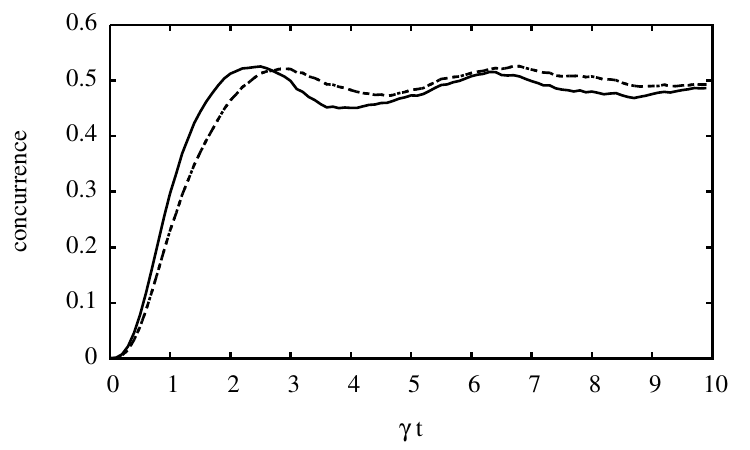}
\par\centering
(c)

\end{minipage}
\hfill
\begin{minipage}[t]{0.49\columnwidth}
\centering
\includegraphics[width=1\textwidth]{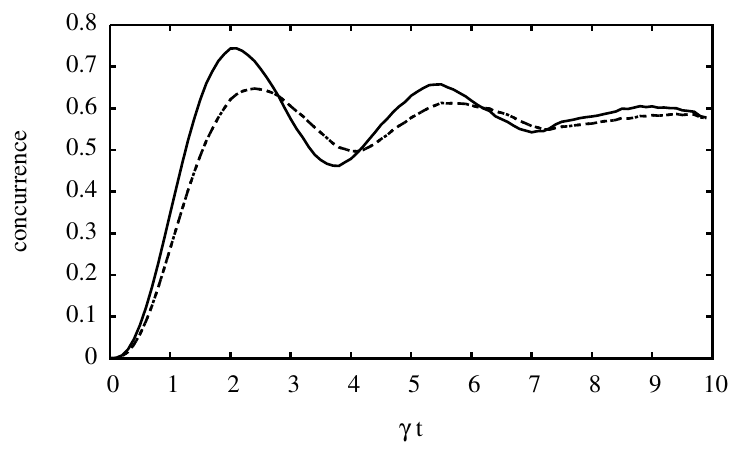}
\par\centering
(d)
\end{minipage}
\caption{A plot of concurrences, $C_{1,23}$ (solid), $C_{2,13}$ (dashed) and $C_{3,12}$ (dotted), calculated without knowledge of which qubit jumps. $Y = 1$.}
\label{fig:2con11}
\end{figure}

In Fig.~\ref{fig:2con11} we plot the same results as in Fig.~\ref{fig:con11}, but this time {\it without} knowledge of which qubit generates the "click" at the detector. Qualitatively the results are similar several important differences.  First, in all four cases the concurrence is roughly twice as large.  We attribute this to the fact that the collective emission can itself generate entanglement whereas the independent emission destroys entanglement.  Second, in Fig.~\ref{fig:2con11}(a) the concurrence is nonzero for all three bipartite splits.  This is because, even though only the second atom is not driven directly, the collective emission will actually cause that atoms to become excited through an indirect process and therefore there can be entanglement between the second atom and the others.  Finally, in Fig.~\ref{fig:2con11}(d) the oscillations are no longer damped out within two lifetimes.  We attribute this to the fact that the dipole-dipole coupling will actually transfer some population into the asymmetric subspace which is completely undamped by the collective emission.

\begin{figure}
\hfill
\begin{minipage}[t]{\columnwidth}
\centering
\includegraphics[width=1\textwidth]{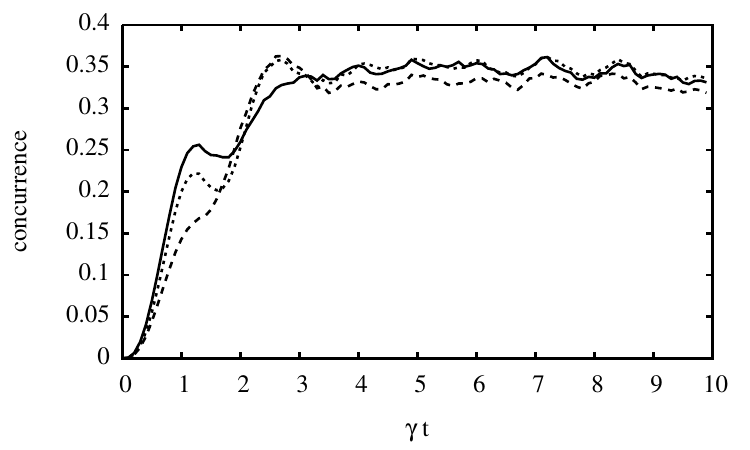}
\par\centering
(f)
\end{minipage}
\caption{A plot of concurrences, $C_{1,23}$ (solid), $C_{2,13}$ (dashed) and $C_{3,12}$ (dotted), calculated with knowledge of which qubit jumps and $g_{23} \rightarrow \imath g$. $Y = 1$.}
\label{fig:3con11}
\end{figure}

In Fig.~\ref{fig:3con11}, we change the phase of $g_{23}$ by $\pi/2$. The only case that changes significantly is with all three couplings nonzero and equal in magnitude.   The phase change results in oscillations in $C_{3,12}$ and $C_{2,13}$ as well as a splitting of those two, which were previously equal.  In the Jaynes-Cummings model, one can use a coupling of $g(\sigma_-+\sigma_+)$ or $ig({\sigma_+-\sigma_-})$ where the $i$ can be absorbed into a phase. However with two couplings, where we use one of each, and the change in relative phase matters.

\section{Residual Entanglement (The Three-Tangle)}
When these three qubits are entangled with each other, not any two of them can be fully entangled with one another unlike classical correlations where it can be shared freely. Qubit 1's entanglement with qubit 2 and its entanglement with qubit 3 is limited and is given by the inequality \cite{3tangle}
\begin{equation}
\label{Eq:ineq}
C^{2}_{12} + C^{2}_{13} \leq C^{2}_{1,23}
\end{equation}
where $C_{12}$ is the concurrence of the mixed state of qubits 1 and 2 given by equation \ref{eq:concmixed}, and similarly $C_{13}$ is of qubits 1 and 3. The equality is satisfied if we consider the W state
\begin{equation}
\label{eq:equal}
|\psi\rangle = \alpha|100\rangle + \beta|010\rangle + \gamma|001\rangle
\end{equation}
where we can find $C_{12} = 2|\alpha\beta|, C_{13} = 2|\alpha\gamma|$, and $C_{1,23} = 2|\alpha\sqrt{|\beta|^2 + |\gamma|^2}|$.
The residual entanglement, $\tau_{123}$, is expressed by 
\begin{equation}
\label{Eq:resient}
C^{2}_{1,23} - C^{2}_{12} - C^{2}_{13} = 4|d_1 - 2d_2 +4d_3| = \tau_{123}
\end{equation}
where 
\begin{eqnarray}
\label{eq:threetangle}
d_1 &=& C^{2}_{000}C^{2}_{111} + C^{2}_{001}C^{2}_{110} + C^{2}_{010}C^{2}_{101} + C^{2}_{100}C^{2}_{011}\\
d_2 &=& C_{000}C_{111}C_{011}C_{100} + C_{000}C_{111}C_{101}C_{010} + C_{000}C_{111}C_{110}C_{001}\\
     & & + C_{011}C_{100}C_{101}C_{010} + C_{011}C_{100}C_{110}C_{001} + C_{101}C_{010}C_{110}C_{001}\\
d_3 &=& C_{000}C_{110}C_{101}C_{011} + C_{111}C_{001}C_{010}C_{100}
\end{eqnarray}
This residual entanglement is invariant under permutation of the three qubits. We can consider, for example the Greenberger-Horne-Zeilinger state $(1/\sqrt{2})(|000\rangle + |111\rangle)$ \cite{ghz}. 
One can easily show that for this state, $\tau_{123} = 1$ and for the W \cite{W} state, Eq.~\ref{eq:equal}, $\tau_{123} = 0$.

\begin{figure}
	\centering
		\includegraphics [width = 5in]{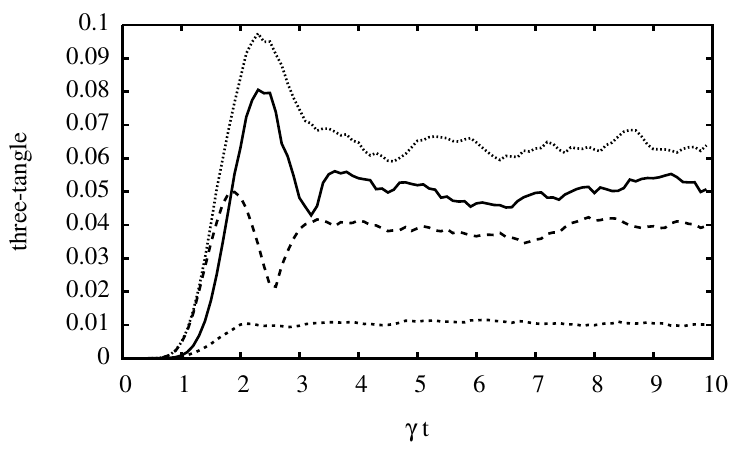}
	 \caption{The three-tangle calculated with knowledge of which qubit jumps.$Y = 1; \gamma_1 = \gamma_2 = \gamma_3 = 1$.}
	 \label{fig:tangle}
\end{figure}

\begin{figure}
	\centering
		\includegraphics [width = 5in]{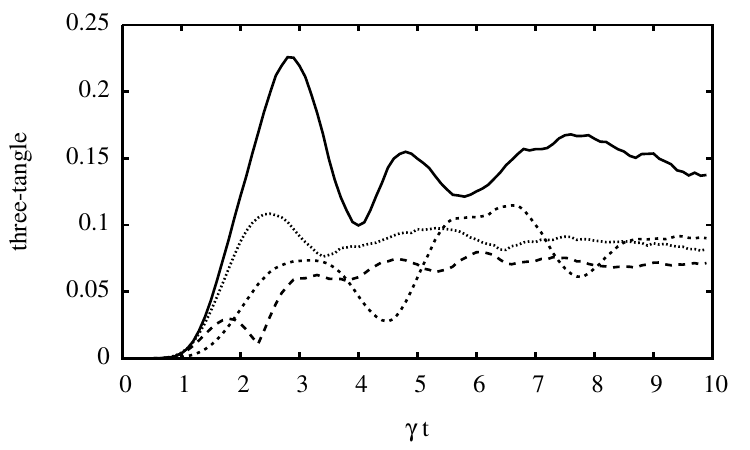}
	 \caption{The three-tangle calculated without knowledge of which qubit jumps. $Y = 1; \gamma_1 = \gamma_2 = \gamma_3 = 1$.}
	 \label{fig:33tangle}
\end{figure}

\begin{figure}
	\centering
		\includegraphics [width = 5in]{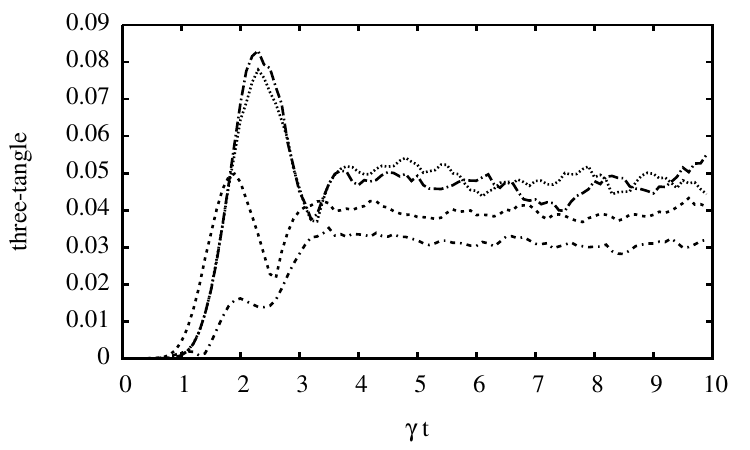}
	 \caption{The three-tangle calculated with knowledge of which qubit jumps and with substitution $g \rightarrow \imath g$. $Y = 1; \gamma_1 = \gamma_2 = \gamma_3 = 1$.}
	 \label{fig:23tangle}
\end{figure}

For the three-qubit state given by equation
 we measured $\tau_{123}$ using the direct detection method averaging over many trajectories as shown in figures \ref{fig:tangle}, \ref{fig:23tangle} and \ref{fig:33tangle}. All results are consistent with the configuration of the qubits. For instance, the residual entanglement is zero for the bipartite configurations and varies for tripartite configurations according to the values of all the other parameters.

In Fig.~\ref{fig:tangle} we calculate the three-tangle without knowledge of which qubit collapses. We find that the three-tangle is zero when only two qubits are coupled, The largest three-tangle is when either $g_{12}$ or $g_{13}$=0, but it is rather small, more like a W state. In the case of all three couplings being nonzero, or no coupling between the undriven qubits, the three tangle is lower by a factor of 10. In Fig.~\ref{fig:33tangle} we plot the same functions {\it with} knowledge of which qubit has collapsed. Qualitatively they are all the same with the exception of the equal coupling case, where we see larger entanglement. In Fig.~\ref{fig:23tangle} we plot the same functions, but with a $\pi/2$ phase shift in the coupling constant between the undriven qubits, the only change is again in the case with three nonzero couplings.

\section{Adiabatic  Elimination}
Here we investigate the results of increasing the decay rate of one of the undriven qubits with respect to the others. We find that the steady state entanglement between qubits 1 and 2 decreases to a minimum value that does not further decrease after the large damping rate is 10 times the other two. However the value of this entanglement is about 40\% less than if the third qubit was not there.  Qubit three can be used as an information port, carrying off information about the other two, but the cost of a large decay rate for it so that we retain basically a bipartite system, there is a loss of entanglement. This will affect protocols where an ancillary qubit is used as a readout port. The entanglement of qubit 3 with the other two decreases to zero, as it should.

\begin{figure}
\centering
\includegraphics [width = 5in]{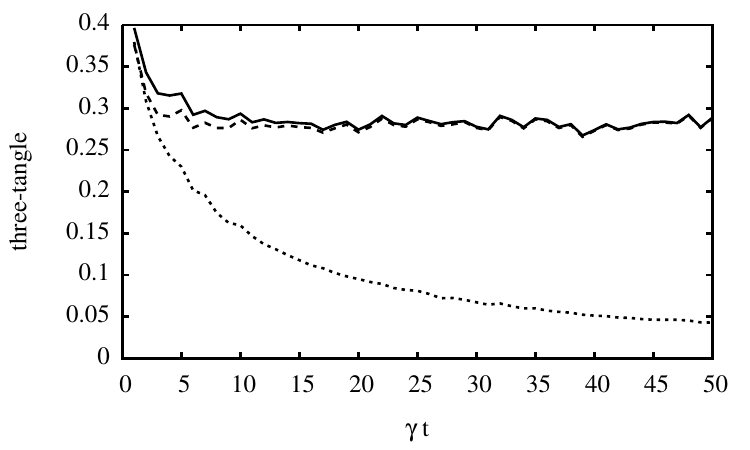}
\caption{Third qubit decays adiabatically carrying off some information which is absent in the mixed state of the other two qubits.$Y = 1;g_{12} = g_{23} = g_{13} = 1;\gamma_1 = \gamma_2 = 1$.}
\label{fig:adiabelm}
\end{figure}

\section{Conclusion}
For a given position, we calculate the dipole-dipole coupling strengths between the atoms in order to obtain the Hamiltonian, with the addition of driving on atom 1.
\\ \indent We also calculate the $\gamma_{ij}$s to generate the jump operators of the system. Using this Hamiltonian and these jump operators, a Monte-Carlo simulation is run from a pure state.
\\ \indent We calculate the average concurrence of the states generated in the quantum trajectories shown in the Monte-Carlo simulations. The concurrence in a three qubit system originates from two sources: the bipartite dipole-dipole coupling and the collective spontaneous emission. We find that knowledge of which qubit emits reduces the entanglement, as that puts the overall state into a product state of one qubit and the other two. We find that using the third qubit as a readout, reduces the available bipartite entanglement available in the other two, which can have impact on applications.

\end{document}